\documentclass[twocolumn,showpacs,preprintnumbers,amsmath,amssymb]{revtex4}
\usepackage{tabularx,graphicx}

\usepackage{color}
\usepackage{hyperref}
\hypersetup{
    colorlinks=true,
    linkcolor=blue,
    filecolor=blue,      
    urlcolor=blue,
}

\usepackage{color}

\usepackage{ulem}   

\begin{document}

\newcommand{\beq}{\begin{equation}}
\newcommand{\eeq}{\end{equation}}
\newcommand{\beqn}{\begin{eqnarray}}
\newcommand{\eeqn}{\end{eqnarray}}
\newcommand{\bmath}{\begin{subequations}}
\newcommand{\emath}{\end{subequations}}
\newcommand{\bra}[1]{\langle #1|}
\newcommand{\ket}[1]{|#1\rangle}

\title{Faulty evidence for superconductivity in  ac magnetic susceptibility of sulfur hydride under pressure}

\author{J. E. Hirsch }
\address{Department of Physics, University of California, San Diego,
La Jolla, CA 92093-0319}

\begin{abstract} 
It is generally believed that  sulfur hydride under high pressure is 
a high temperature superconductor \cite{sh3,picketteremets,review2,shimizu,boosted,last,progress}.
In National Science Review 6, 713 (2019) Huang and coworkers 
 Ref. \cite{huang} reported detection of superconductivity in sulfur hydride through a highly sensitive ac magnetic susceptibility technique and an unambiguous
determination of the superconducting phase
diagram.
In this paper we present evidence  showing that the experimental results reported in that paper\cite{huang} 
do not support the conclusion that sulfur hydride is a superconductor.
\end{abstract}
\pacs{}
\maketitle 

\section{introduction}

In 2015, Drozdov and coworkers reported the discovery of high temperature superconductivity in sulfur hydride \cite{sh3}.
The result is generally assumed to be true \cite{picketteremets,review2,shimizu,boosted,last,progress}. 
Independently, Huang and coworkers measured ac magnetic susceptibility of sulfur hydride \cite{huang}
and in appearance confirmed the existence of superconductivity. According to ref.   \cite{semenok},
that work   ``sets a new standard for experimental
studies of superconductivity at high
pressure''.

However, we have recently argued that the experimental  evidence supporting superconductivity in sulfur hydride
presented in ref. \cite{sh3} is not convincing 
\cite{hm4}, and that neither is the one presented in refs. \cite{nrs,nrs2} concerning the Meissner effect  \cite{hm3,hm5}.  In this paper  we argue  that the ac susceptibility measurements of ref. \cite{huang} 
provide no support for the existence of superconductivity in sulfur hydride either. No other studies of ac magnetic susceptibility nor of other
magnetic properties of sulfur hydride besides those mentioned above
have been reported so far.

Ac magnetic susceptibility is a superior test  for superconductivity in 
materials under high pressure \cite{klotz,tim,back2,hamlinthesis,md,yb}. 
A superconductor excludes magnetic flux, so upon cooling into the superconducting state
a sharp drop in the ac magnetic susceptibility is observed.
For experiments under high pressures, because of the smallness of the sample required by the 
geometry of the diamond anvil cell, the detected signal is   a tiny drop in a large signal 
arising from the superposition of the sample and the background magnetic responses, with the background signal being several
orders of magnitude larger than the sample signal \cite{klotz,tim,hamlinthesis,yb}. 
For that reason, it is customary to subtract from the total signal (the so-called "raw data")  the background signal,
according to the relation
\beq
data = raw \;   data - background \;  signal.
\eeq
The background signal is  usually obtained by measuring the susceptibility at a pressure value such that no superconducting 
transition occurs in the temperature range of interest  \cite{hamlinthesis}.

Reference \cite{huang}    reports measurements of ac magnetic susceptibility measurements for sulfur hydride at seven different pressure values.
It only shows results for ac susceptibility for four values of pressure.  These results (figure 2 of ref. \cite{huang}) are reproduced in Fig. 1.
The drops in the signals seen in the figure were interpreted as due to the onset of superconductivity
at the critical temperatures $T_c$  shown in the figure \cite{huang}.

\begin{figure}[h]
 \resizebox{8.5cm}{!}{\includegraphics[width=6cm]{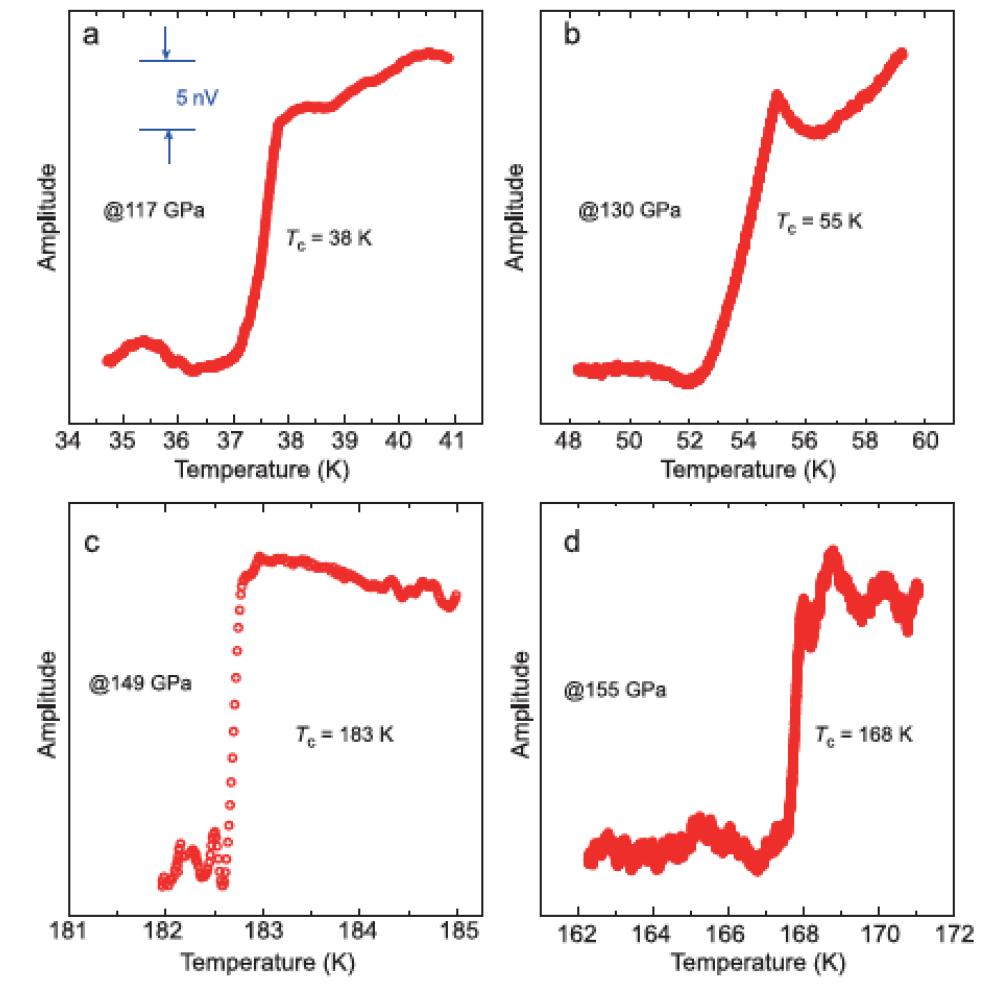}} 
\caption{Magnetic susceptibility signals of sulfur hydride at various pressures, Fig. 2 of ref. \cite{huang}.
}
\label{fig3}
\end{figure}
 
 \section{The measured   raw data}
The results shown in Fig. 1 are for the left side of Eq. (1), i.e. are obtained by  subtracting a background signal from raw data.
In order to assess the validity and significance of these results, we requested from the authors
the raw data and background signal measured, from which the results shown in Fig. 1 were obtained.
Following generally accepted proper scientific practice, the authors kindly sent us  these data recently   as
well as gave us additional details  of the measurements upon our request \cite{pcs}.

\begin{figure}[]
 \resizebox{8.5cm}{!}{\includegraphics[width=6cm]{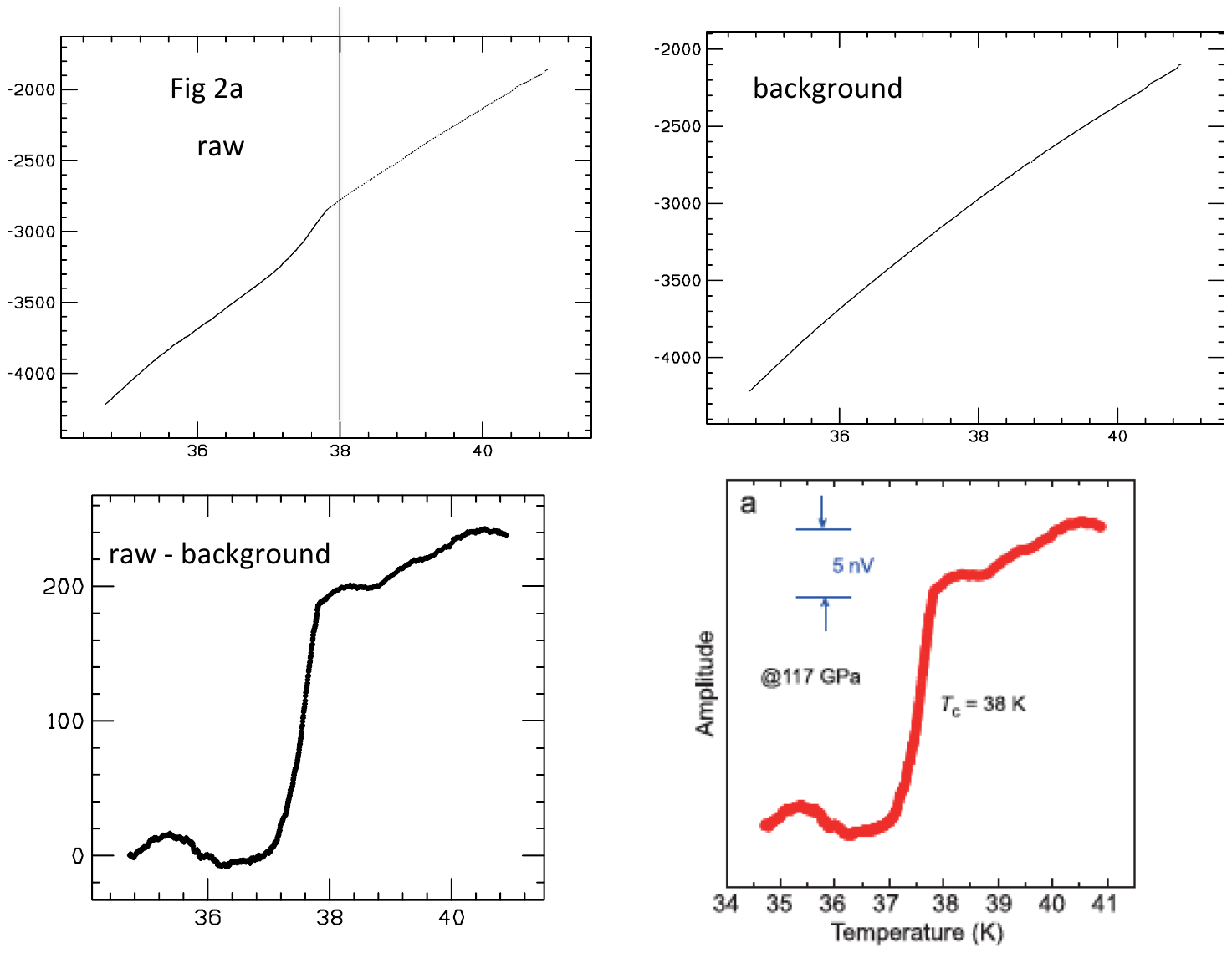}} 
\caption{Top panels: raw data and background signal for Fig. 2a   of ref. \cite{huang} plotted from data supplied by the authors \cite{pcs}.
The lower left panel shows the difference between the upper left and right panels, in agreement with the published
results shown on the lower right panel.
}
\label{fig3}
\end{figure}

\begin{figure}[]
 \resizebox{8.5cm}{!}{\includegraphics[width=6cm]{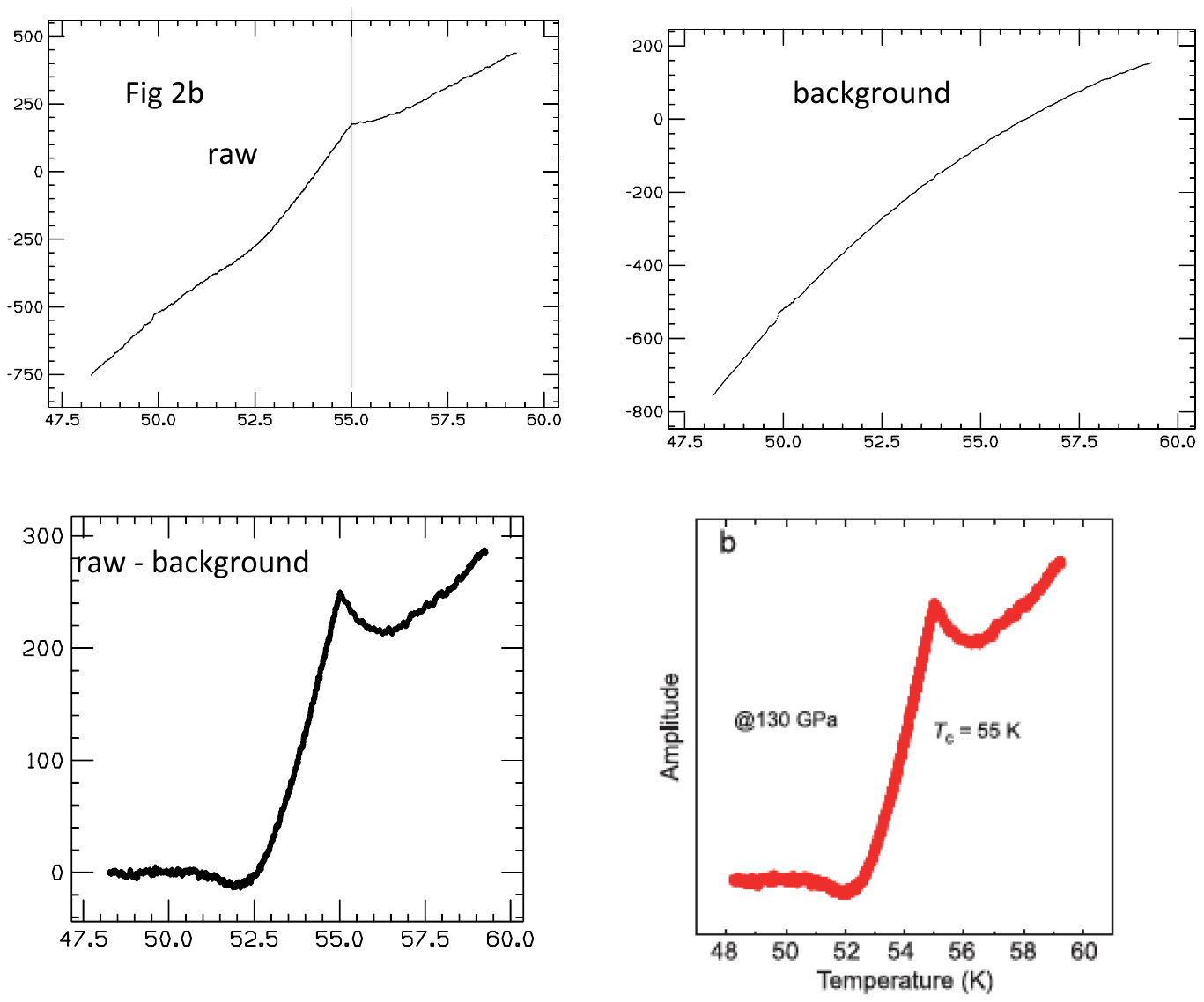}} 
\caption{Top panels: raw data and background signal for Fig. 2b   of ref. \cite{huang} plotted from data supplied by the authors \cite{pcs}.
The lower left panel shows the difference between the upper left and right panels, in agreement with the published
results shown on the lower right panel.
}
\label{fig3}
\end{figure}

\begin{figure}[]
 \resizebox{8.5cm}{!}{\includegraphics[width=6cm]{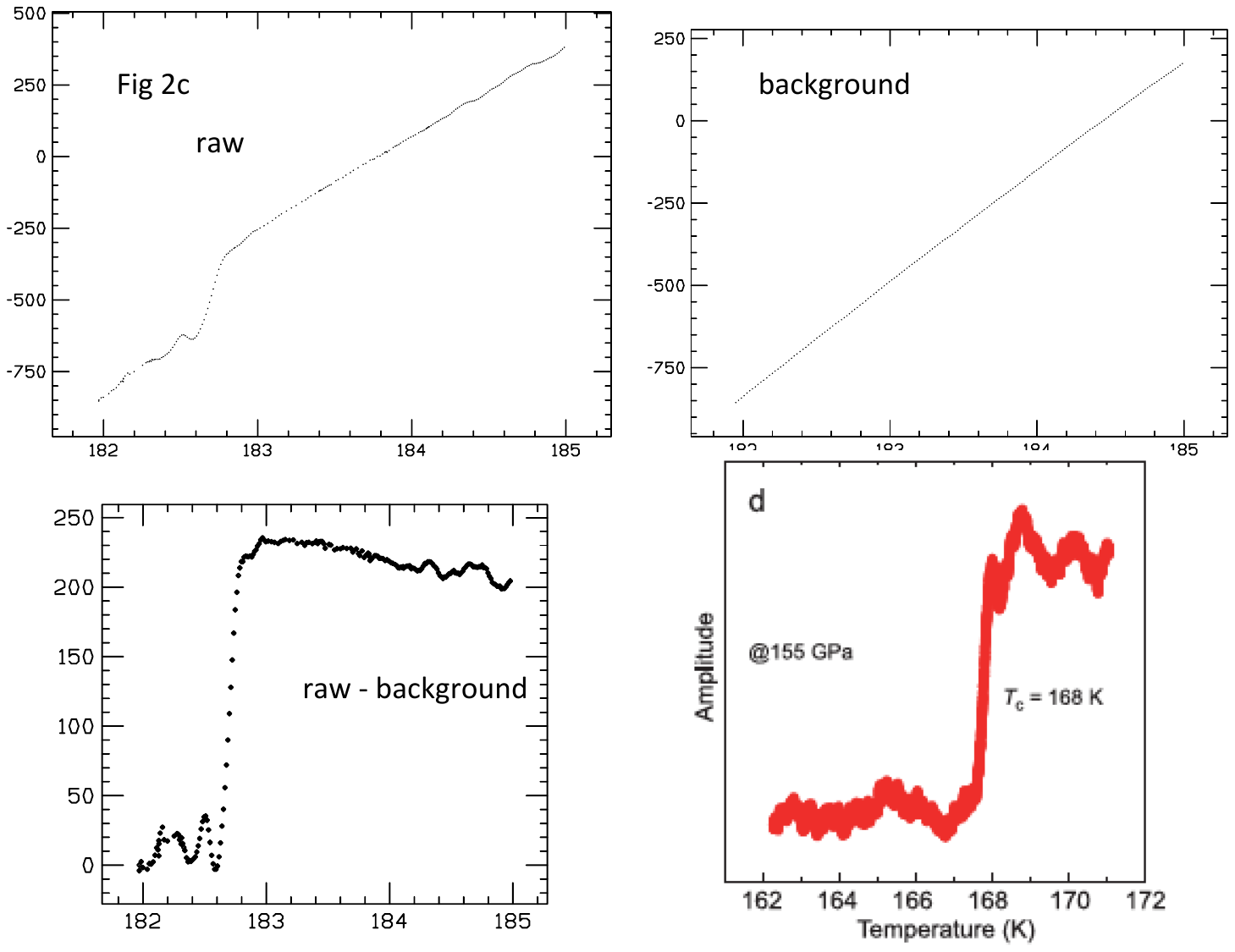}} 
\caption{Top panels: raw data and background signal for Fig. 2c   of ref. \cite{huang} plotted from data supplied by the authors \cite{pcs}.
The lower left panel shows the difference between the upper left and right panels, in agreement with the published
results shown on the lower right panel.
}
\label{fig3}
\end{figure}

\begin{figure}[]
 \resizebox{8.5cm}{!}{\includegraphics[width=6cm]{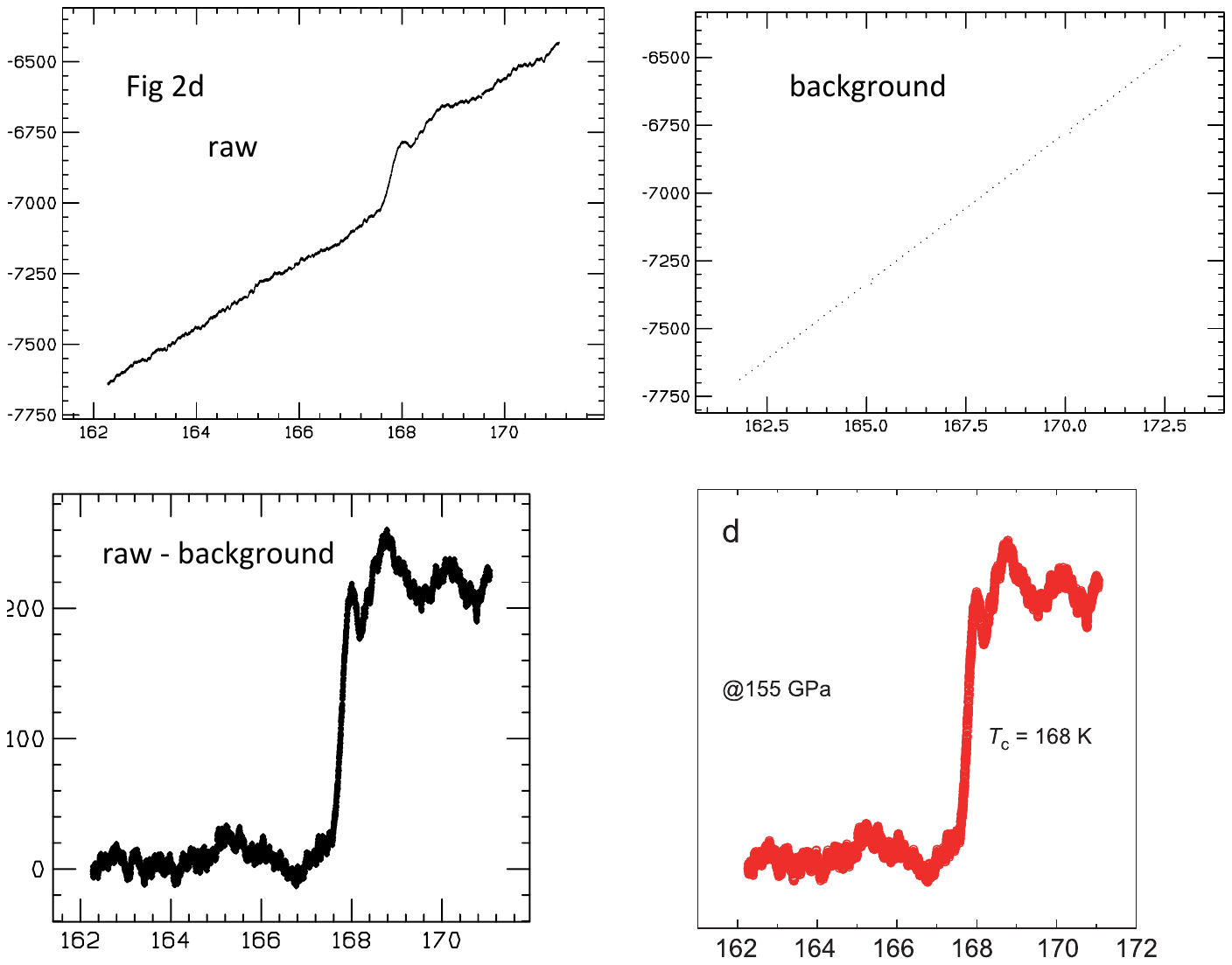}} 
\caption{Top panels: raw data and background signal for Fig. 2d   of ref. \cite{huang} plotted from data supplied by the authors \cite{pcs}.
The lower left panel shows the difference between the upper left and right panels, in agreement with the published
results shown on the lower right panel.
}
\label{fig3}
\end{figure}
 
From the data supplied by the authors, we made plots of the raw data and the background signal and of their difference according to 
Eq. (1). These are shown in Figs. 2 to 5. It can be seen that the difference calculated from the raw data and background signal
supplied by the authors properly reproduces the data plotted in Fig. 2 of ref.  \cite{huang}. 

We also learn from these figures that the background signal, which was collected separately for each temperature range at a lower pressure \cite{pcs}, is smooth in all cases, with
nearly constant slope around the region where the transition occurs. This is what one expects
for these measurements, as also seen for example in 
ref. \cite{yb}. As a consequence, the raw data shown in the top left panels of Figs. 2 to 5 show similar slope versus temperature
above and below $T_c$.

This is in stark contrast with the  raw data for ac magnetic susceptibility of a carbonaceous
sulfur hydride (CSH) under pressure, claimed to be superconducting at room temperature, shown in ref. \cite{roomt}.
For that material, the published raw data for ac magnetic susceptibility (Extended Data Fig. 7 d of ref.  \cite{roomt}) show a very significant change in slope across the transition, similar to what such data look like for the element europium at low temperature \cite{eu}. 
I have suggested that this implies that the published data for ac susceptibility of CSH are 
invalid \cite{eumine,eumine1,eumine2}.
Instead, the authors of ref. \cite{roomt} state that the background signal is responsible for this unusual behavior \cite{diasscience}.
This would imply a very anomalous background signal, qualitatively different from that seen in Figs. 2-5 here, 
arguably impossible  \cite{eumine2}.
The authors of ref. \cite{roomt} have declined to supply their raw data and background signal measurements to verify or disprove their claim.

 \section{further analysis}
 
 \begin{figure}[]
 \resizebox{8.5cm}{!}{\includegraphics[width=6cm]{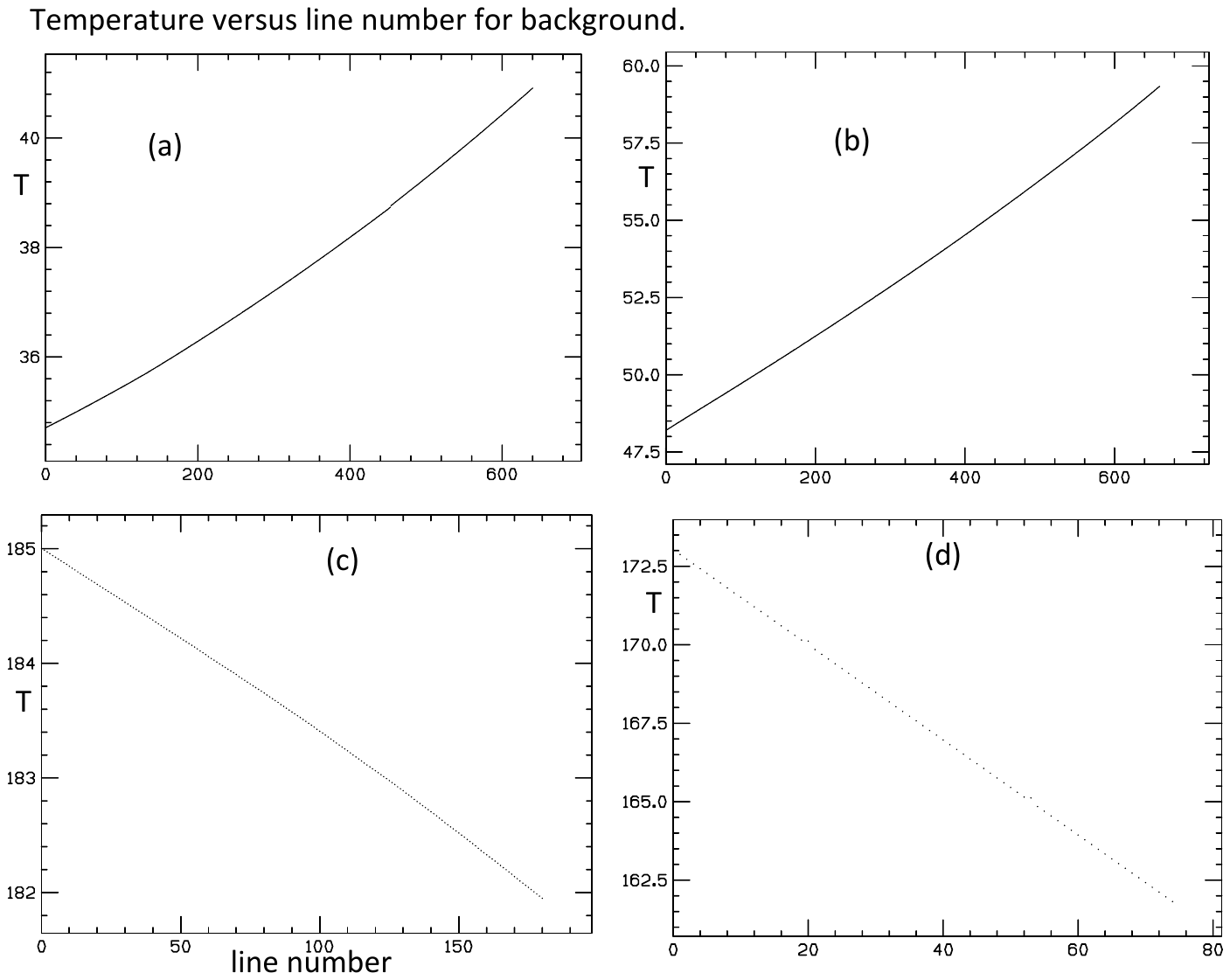}} 
\caption{Temperature versus line number in the file for the background signal 
measurements used to obtain the data in Fig. 1, measured at  pressures lower than the
values where superconducting transitions occur.
}
\label{fig3}
\end{figure}

To further examine the validity and significance of the susceptibility data for sulfur hydride reported in \cite{huang} under discussion here,
we have made graphs of the measured temperature versus line number in the data files supplied
by the authors \cite{pcs}. The authors did not give us a a time track record for each measurement, but informed us that data were taken at a constant rate of temperature change, mostly about
0.1K/min, sometimes no bigger than 0.5K/min \cite{pcs}. With the reasonable assumption that
data were taken at regular time intervals, we would expect a smooth curve for
the temperature as a function of line number on the data file. Indeed, that is what 
we see for the background signal measurements, shown in Fig. 6.

 \begin{figure}[]
 \resizebox{8.5cm}{!}{\includegraphics[width=6cm]{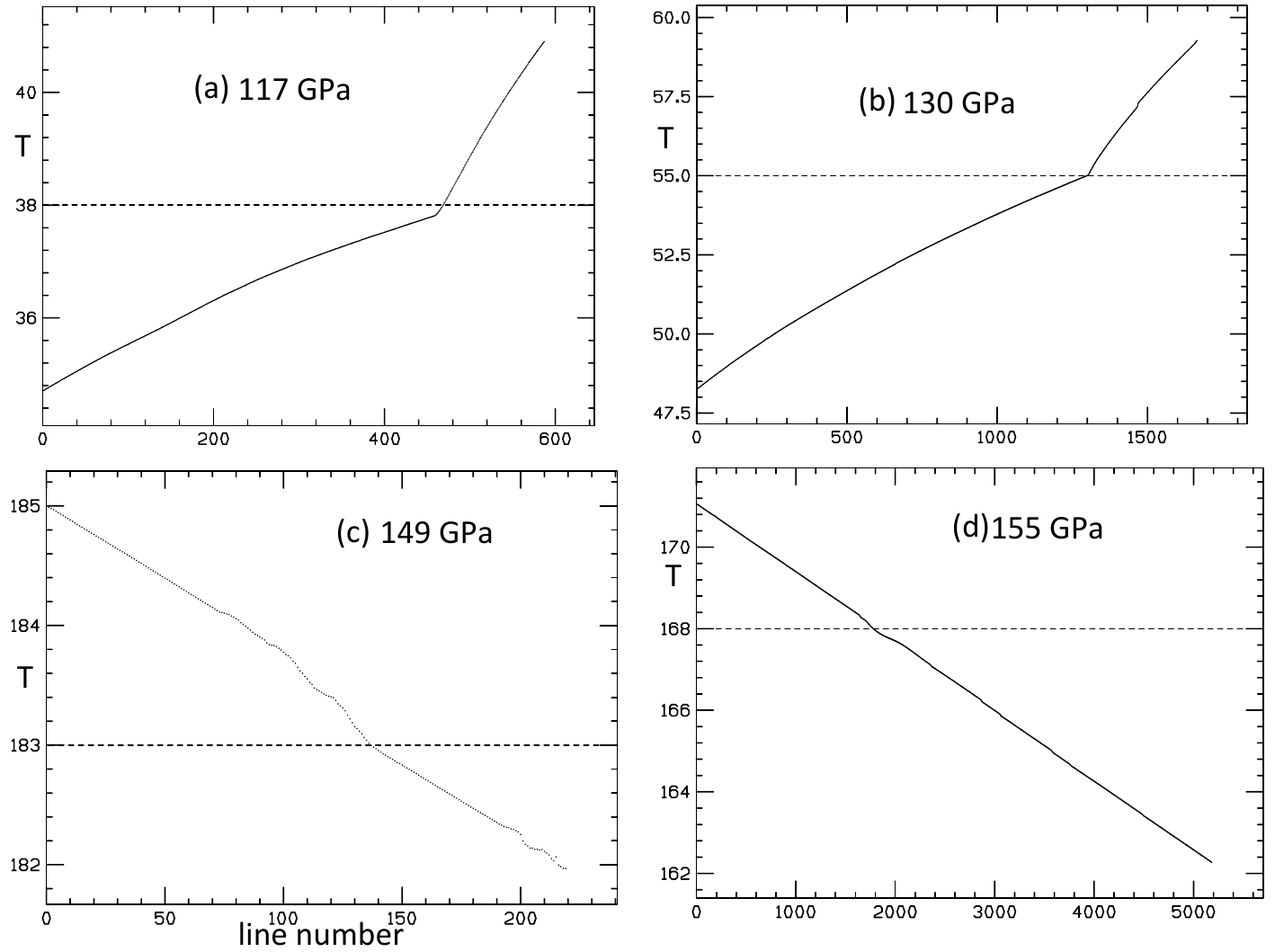}} 
\caption{Temperature versus line number in the file for the raw data 
measurements for the four cases given in Fig. 1. The dashed lines indicate
the superconducting transition temperatures given in Fig. 1.
}
\label{fig3}
\end{figure}

Instead, when we do the same plots for the different temperature ranges shown in Fig. 1 at the pressure
values given in Fig. 1 where the supposed superconducting transitions occur,
we find the very surprising results shown in Fig. 7.
Particularly for cases (a) and (b), a sharp break in the temperature
versus line number in the file occurs right at the point where the  assumed
superconducting transition takes place.

The behavior shown in Fig. 7 is very unexpected. It suggests that the 
rate of temperature change is not constant in time, contrary to what we were told was
the experimental protocol \cite{pcs}. Indeed, in Fig. 8 we plot $\Delta T$, the change in
temperature between two subsequent lines in the data file, versus temperature for two cases.
There is a sharp break at the critical temperature, with the temperature changing
faster above than below $T_c$.

\begin{figure}[]
 \resizebox{8.5cm}{!}{\includegraphics[width=6cm]{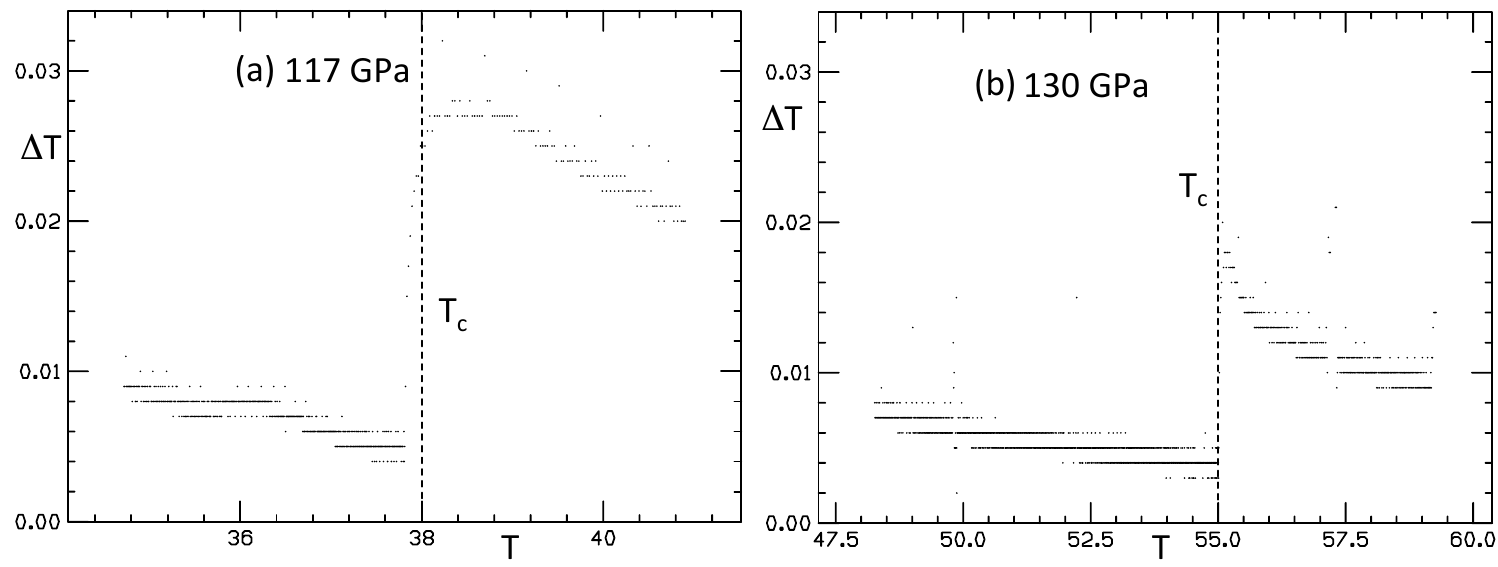}} 
\caption{Change in temperature ($\Delta T$)  between two 
subsequent lines in the data file for the first two cases of Fig. 1, versus temperature.}
\label{fig3}
\end{figure}

If the system is being heated at a constant rate, as we were informed \cite{pcs}, the sudden increase
in the temperature change could indicate that the heat capacity of the system suddenly decreased.
For a superconducting transition this is in fact expected: the specific heat jump at the
critical temperature is given by $(c_s-c_n)/c_n=1.43$, with $c_n$ and $c_s$ the heat
capacities in the normal and superconducting states.

However, the temperature sensor cannot be placed in the diamond anvil cell next to the sample, that is   physically impossible \cite{semenokpc}. Assuming the temperature sensor is located at a distance
$R\sim 1cm$ from the sample, the temperature measured would correspond to that of a volume of order 
$\sim 10^7$ times larger than the volume of the sample, so it cannot possibly be influenced
to the degree shown in Fig. 8  by 
a change in the heat capacity of the sample.

Therefore, we have to conclude that the sudden change in the 
temperature increment coinciding with the assumed critical temperature is an experimental artifact.
This indicates that the observed change in slope in the ac susceptibility observed at those
temperatures is a consequence of the same experimental artifact and cannot be taken as
evidence of a superconducting transition.

Fig. 9 shows the corresponding plots for the other two values of pressure, 149 GPa and 155 GPa.  
There is also  evidence here that there are changes in the rate of temperature change right at the assumed
superconducting transitions, which again is not expected and casts doubt on the conclusion that the
observed drops in Figs. 1c and 1d indicate superconductivity.

\begin{figure}[]
 \resizebox{8.5cm}{!}{\includegraphics[width=6cm]{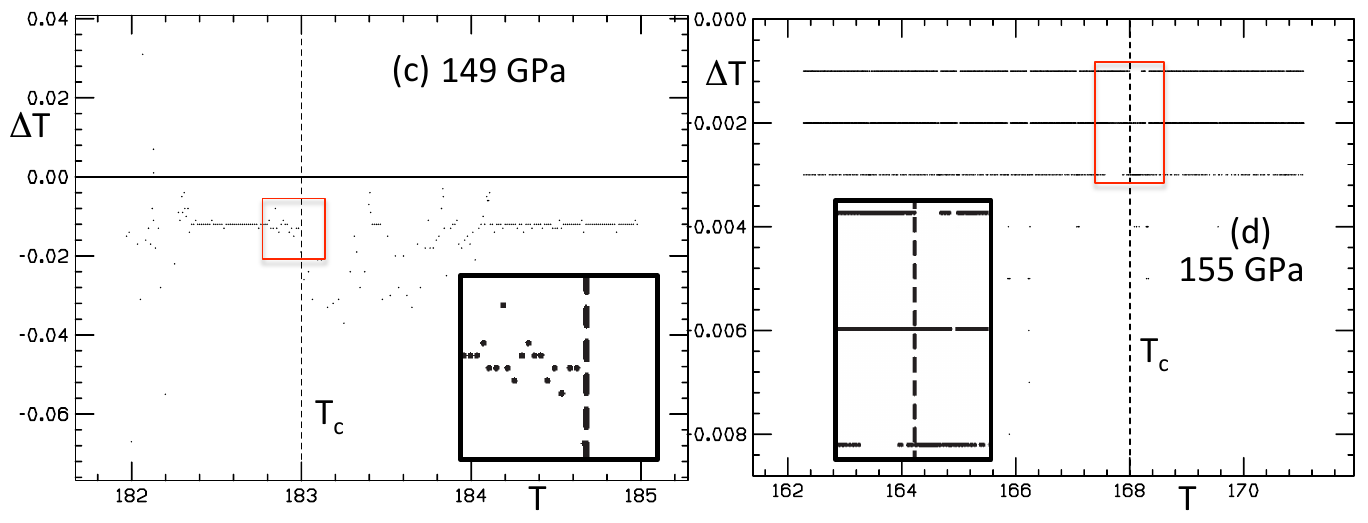}} 
\caption{Change in temperature ($\Delta T$)  between two 
subsequent lines in the data file for the last two cases of Fig. 1, versus temperature.
The insets show the regions enclosed in the red rectangles enlarged, showing the change across $T_c$.
}
\label{fig3}
\end{figure}

\section{Speculation on a possible scenario}

In this section we ask: assuming the sample did not go superconducting at the $T_c$'s indicated in
Fig. 1, what could be a possible scenario that led to the obtention of Fig. 1, which indicates that
there were superconducting transitions when in fact there were none?

Given the anomalous temperature steps found in Fig. 8, we hypothesize that there may also have been
 a problem
with the temperature determination, i.e. that the readings in the thermometer did not accurately reflect
the temperature of the sample. As a simple scenario, let us assume that for 117 GPa the
temperature step at the sample location was in fact constant below $T_c$, given by the average value of what was measured below
$T_c$, $\Delta T=0.0068K$, and jumped up  by a factor of
three at $T_c$, similarly to what the left panel of Fig. 8 shows, remaining constant above $T_c$. This is indicated by
the dashed red lines on the left panel of Fig. 10, which is not too different from what was measured
(black lines). The resulting raw data versus this new temperature scale
are shown on the right panel of Fig. 10, and no longer show the structure near $T_c$ that the
original raw data showed (top left panel of Fig. 2). If we now subtract the background, we obtain
what is shown on the left panel of Fig. 11. The steep susceptibility drop at a temperature interpreted as the
superconducting transition temperature has disappeared.

\begin{figure}[]
 \resizebox{8.5cm}{!}{\includegraphics[width=6cm]{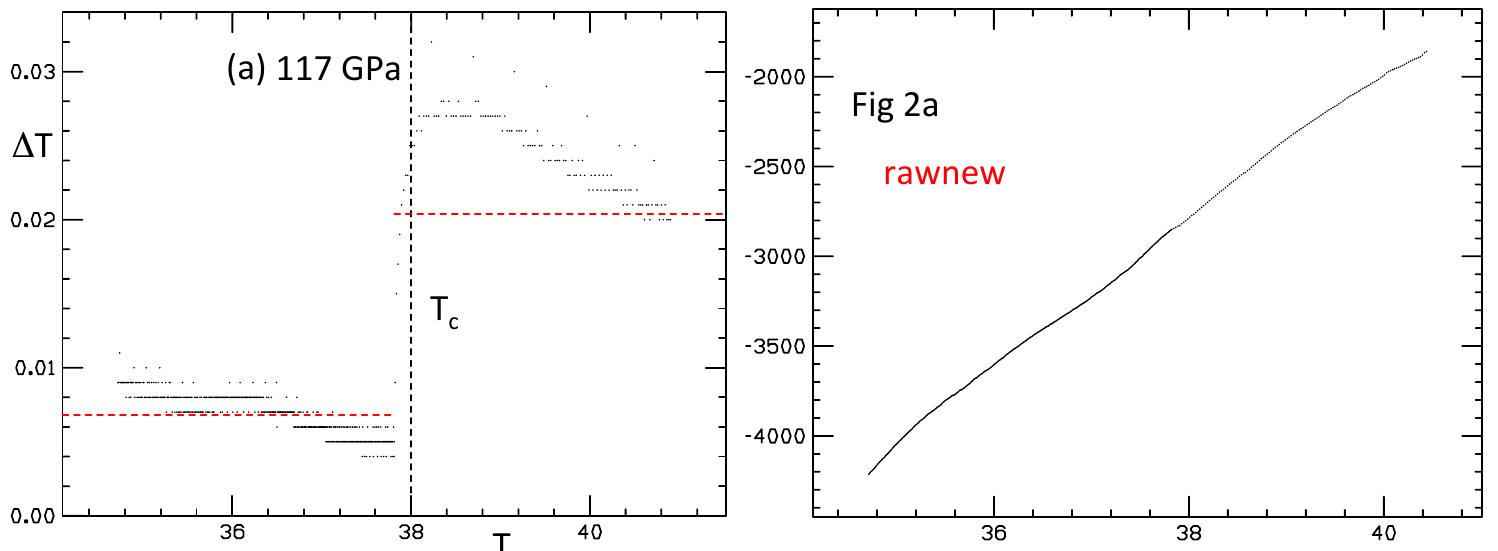}} 
\caption{Left panel: we assume that the temperature steps were in reality given by the red dashed lines, constant below
and above $T_c$, changing by a factor of 3 at $T_c$. The right panel shows the resulting raw data plotted
versus this new temperature scale. }
\label{fig3}
\end{figure}

\begin{figure}[]
 \resizebox{8.5cm}{!}{\includegraphics[width=6cm]{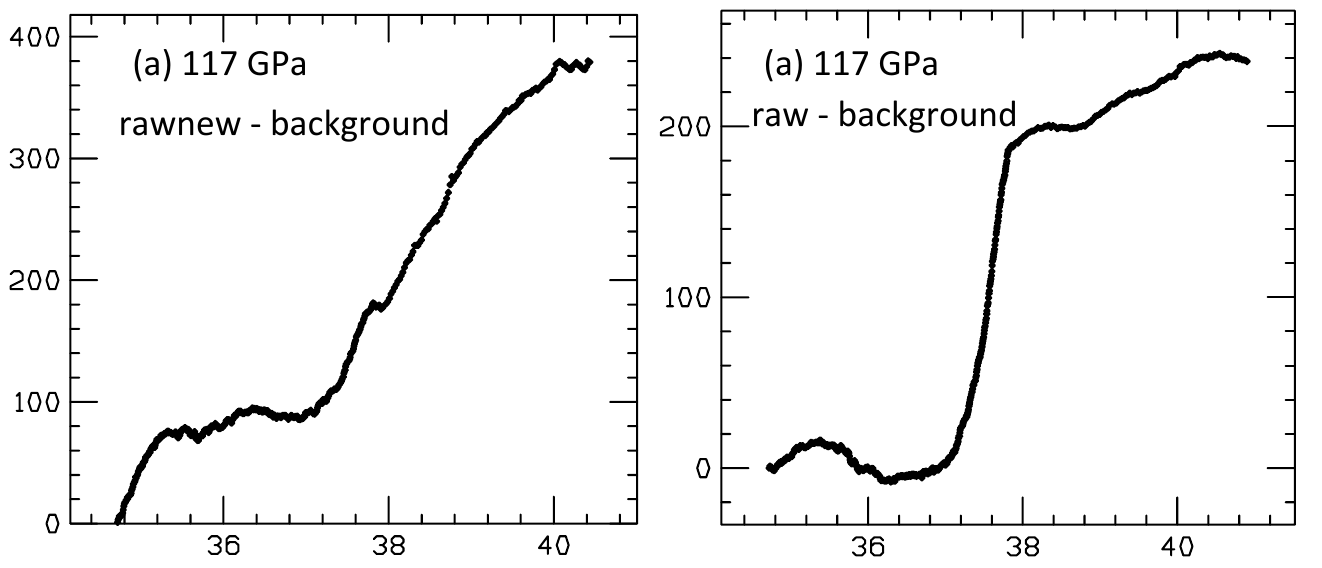}} 
\caption{The left panel shows the susceptibility after background subtraction assuming the
raw data given in the right panel of Fig. 10, i.e. with the new temperature scale. Note that the jump reported to occur around
38 K, shown in the right panel, is no longer there.
}
\label{fig3}
\end{figure}

\begin{figure}[]
 \resizebox{8.5cm}{!}{\includegraphics[width=6cm]{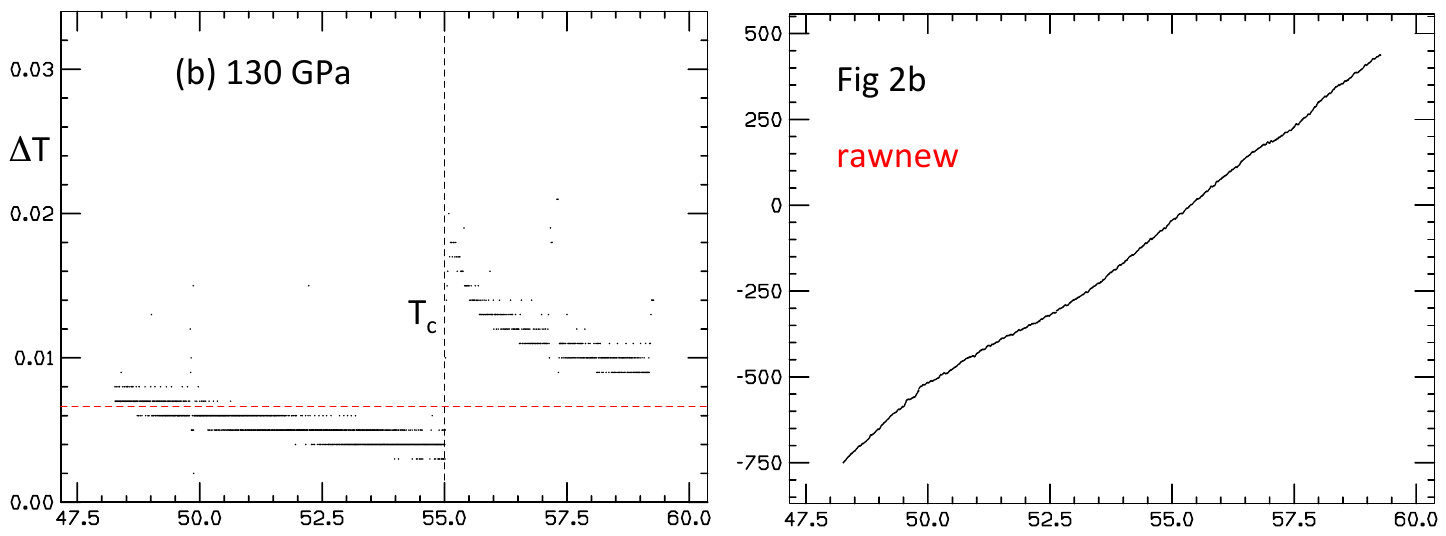}} 
\caption{Left panel: we assume that the temperature steps were in reality given by the red dashed line, i.e. the
same for all temperatures. The right panel shows the resulting raw data plotted
versus this new temperature scale. }
\label{fig3}
\end{figure}

\begin{figure}[]
 \resizebox{8.5cm}{!}{\includegraphics[width=6cm]{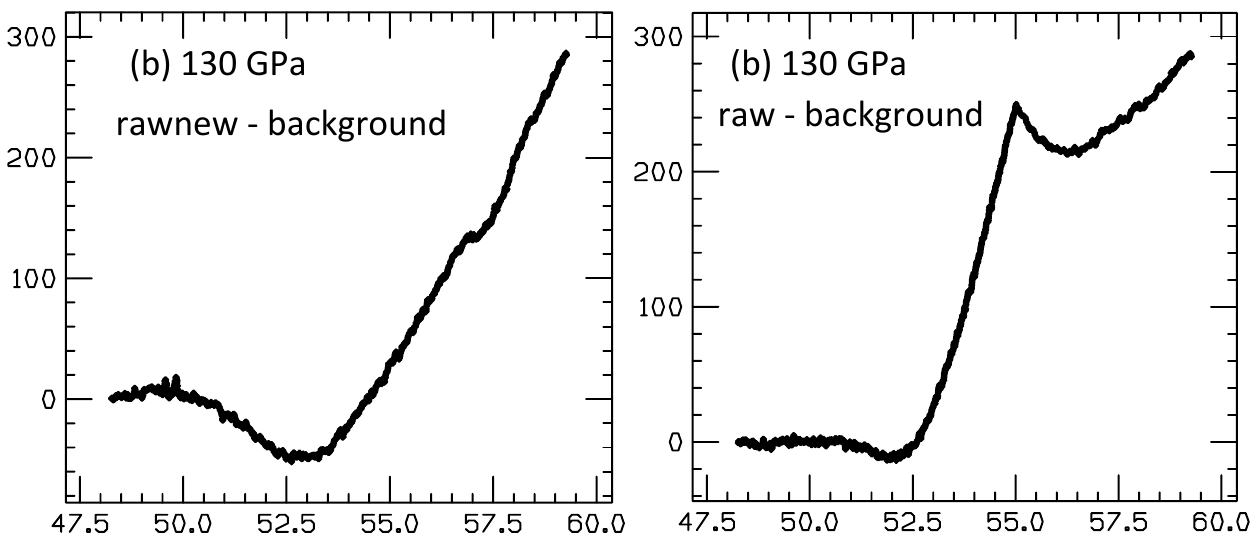}} 
\caption{The left panel shows the susceptibility after background subtraction assuming the
raw data given in the right panel of Fig. 12, i.e. with the new temperature scale. Note that the jump reported to occur around
55 K, shown in the right panel, is no longer there.
}
\label{fig3}
\end{figure}

For 130 GPa, we assume again that the step was constant below  $T_c$ and above $T_c$, but here we will
assume that the value was the same below and above $T_c$, given by the average of the step value
over the entire temperature range where data were given, $\Delta T=0.0066K$, shown by the red dashed line in the left panel of Fig. 12.  The resulting raw data versus this new temperature scale
are shown on the right panel of Fig. 12, and again no longer show the structure near $T_c$ that the
original raw data showed (top left panel of Fig. 3). If we now subtract the background, we obtain
what is shown on the left panel of Fig. 13. There is no structure around the original $T_c\sim 55K$, both the
rise above $T_c$ and the steep drop below $T_c$ seen in the right panel of Fig. 13 are gone.

We conclude from this analysis that in the experiment there may have been some unwanted/uncontrolled
variations in the temperature step, as well as some discrepancies between what was the actual value
of the temperature at the sample position versus what the thermometer measured at a different
position. This, combined with a bias towards taking seriously results that seemed to agree
with expectations, and discarding other results that did not, may have led to ending up with
the results shown in Fig. 1 that did not properly reflect the physics going on in the sample.

\section{conclusions}

In this paper we have analyzed the raw data \cite{pcs} associated with the ac magnetic susceptibility measurements
of sulfur hydride \cite{huang} that were interpreted as providing evidence that sulfur hydride under pressure
becomes superconducting \cite{huang,semenok}, supporting the claim of ref. \cite{sh3}.
We concluded that the published data were correctly inferred from obtained  measurements of
raw data and background signal by subtraction, as given by Eq. (1).

However, when considering the rate of change of temperature of the system
shown in the raw data for the cases where a superconducting transition was supposedly detected, we found
an anomalous jump that exactly lines up with the inferred transition temperature. This,
and the fact that the anomaly is not found in the background signal measurements, strongly suggests
that the observed change in slope of the susceptibility curves is not an indication of a 
superconducting transition but instead is   an experimental artifact. 

Furthermore, we hypothesized that as a consequence of this experimental problem the measurements of the
temperature were not necessarily accurate, and with  small modifications of the temperature scale we showed
for two cases that the jumps interpreted as indicating a superconducting transition completely disappear.

It should also be noted that the rise in the amplitude of the signal right before the assumed
superconducting transition as the temperature is lowered  seen
in Fig. 1b, for 130 GPa, is anomalous. The drop in susceptibility due to 
the onset of superconductivity should not be preceded by a susceptibility rise.
In the raw data, this is signaled by the flattening of the curve seen in the upper left panel
of Fig. 3. This fact alone should cast doubt on the validity of the results reported in Fig. 1b, even in the
absence of the other anomalies pointed out in this paper.

Ac magnetic susceptibility measurements of sulfur hydride have not been independently reproduced.
In view of the results presented here, we urge that these experiments be repeated to 
establish whether or not they provide support for the claimed superconductivity of
sulfur hydride \cite{sh3}. We predict they will not. Nor has other claimed magnetic evidence for 
superconductivity of sulfur hydride \cite{sh3,nrs}  been independently reproduced, 
and its validity has been questioned \cite{hm4,hm3,hm5}.

\begin{acknowledgments}
I am grateful to the authors of ref. \cite{huang} and in particular to Dr. 
Xiaoli Huang for sending me the raw data and additional information on the measurements.
Additionally, I am grateful to S. K. Sinha, A. Fra\~{n}o and M. B. M. Maple for helpful discussions.

\end{acknowledgments}

\appendix 
  \section{Additional information}
  A shortened version of this paper   \cite{huangmine} (due to length constraints in the journal) was published  in Nat. Sci. Rev.  as a Comment  to Ref. \cite{huang}. In a Reply \cite{huangreply},
  the authors of Ref. \cite{huang} argued  that {\it ``our measured data demonstrate that there are
no relationships between the superconducting transition signals and those temperature breaks.''}
  
  To address this point, we show in Fig. 14 the measured raw susceptibility data and the temperature increments, with high resolution, for two values
  of the pressure.
  It can be seen in Fig. 14 that the temperature values where the susceptibility sharply changes its slope and where the temperature increments sharply
  change, coincide to better than 1 part in 1000 for both pressure values. We argue that 
  this indicates that with very high probability there $is$ a relationship between the superconducting transition signals and the 
  temperature breaks, contrary to what the authors say in their Reply \cite{huangreply}.
  
  \begin{figure*}[]
 \resizebox{18.5cm}{!}{\includegraphics[width=6cm]{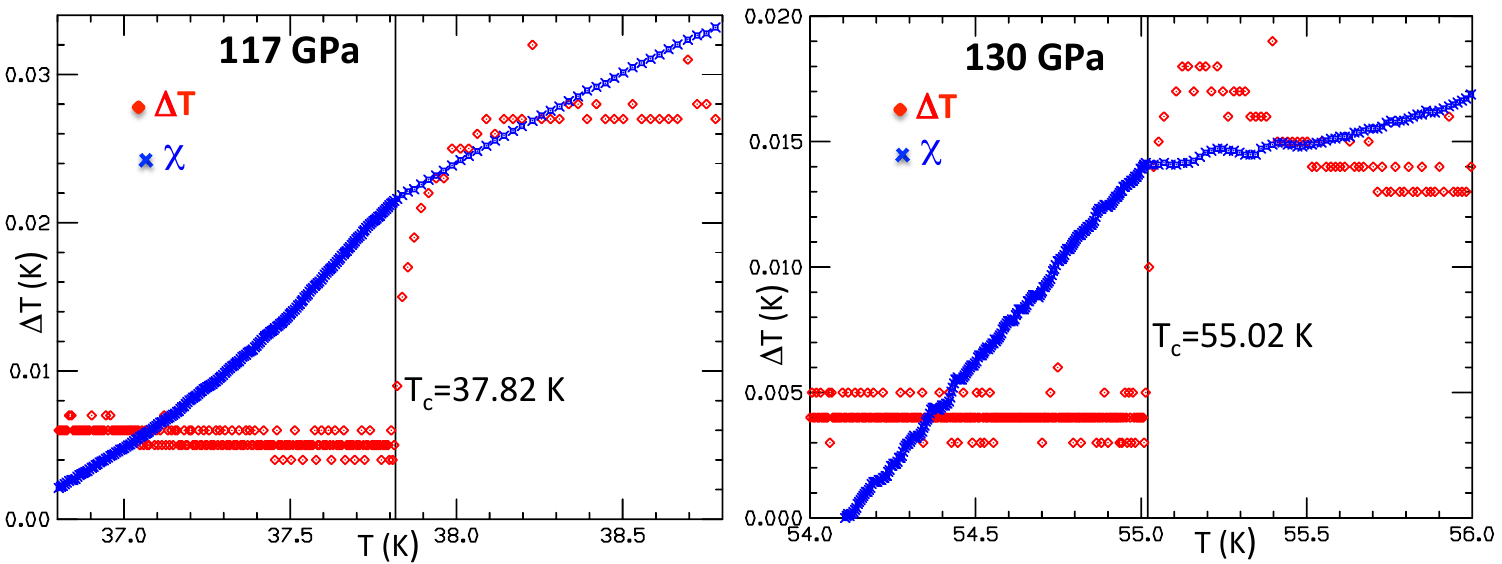}} 
\caption{Comparison of the position of the breaks in the temperature increments $\Delta T$ and the kinks in the raw susceptibility data
for pressure values 117 GPa and 130 GPa.}
\label{fig3}
\end{figure*}

\end{document}